\documentclass[twocolumn,showpacs,preprintnumbers,amsmath,amssymb]{revtex4}

\usepackage{graphicx}
\usepackage{dcolumn}
\usepackage{bm}
\usepackage[caption=false]{subfig}
\usepackage{epsfig}
\usepackage{graphics}
\usepackage{graphicx}
\usepackage{mathtext}

\begin{document}

\title{New universality class in percolation on multifractal scale-free planar stochastic lattice}

\author{M. K. Hassan and M. M. Rahman}
\affiliation{University of Dhaka, Department of Physics, Theoretical Physics Group, Dhaka-1000,
Bangladesh.}

\begin{abstract}%
We investigate site percolation on a weighted planar stochastic lattice (WPSL) which is a multifractal
and whose dual is a scale-free network. Percolation is typically characterized by percolation threshold $p_c$ and by
a set of critical exponents $\beta$, $\gamma$, $\nu$ which describe 
the critical behavior of percolation probability $P(p)\sim (p_c-p)^\beta$, mean cluster size $S\sim (p_c-p)^{-\gamma}$ and the 
correlation length $\xi\sim (p_c-p)^{-\nu}$. Besides, the exponent $\tau$ characterizes the cluster size distribution  
function $n_s(p_c)\sim s^{-\tau}$ and the fractal dimension $d_f$ the spanning cluster. We obtain an exact value for $p_c$ and  
for all these exponents. Our results suggest that the percolation on WPSL belong to a new universality class as its exponents do not share the same value as for all the existing planar lattices.   

\end{abstract}


\maketitle

Percolation is perhaps one of the most studied problems in statistical physics because it provides 
a general framework of statistical theories that deal with structural and transport properties
in porous or heterogeneous media \citep{ref.Stauffer,ref.bunde}.
To study percolation one has to first choose a skeleton, namely an empty lattice or a graph. 
The model can then be defined by one sentence.
Each site/bond of the lattice or graph is either occupied with probability $p$ or 
remains empty with probability $1-p$  independent of the state of its neighbors.  
For small values of $p$ we see mostly single or a few contiguous occupied sites which are called clusters. 
As $p$ increases, the mean cluster size always keeps growing at an increasingly faster rate till it comes to a state when suddenly a 
macroscopic cluster appears for the first time spanning from one end of the lattice to its opposite end. This sudden
onset of a spanning cluster in an infinite system occurs at a particular value of $p$ known as the percolation threshold $p_c$. 
This is accompanied by 
sudden or abrupt change in the behavior of the observable quantities with a small change in its control parameter $p$.
Such change is almost always found to be the signature of phase transition that occur in a wide range of phenomena
\citep{ref.Stanley, ref.Binney}. This is 
why scientists in general and physicists in particular find percolation theory so attractive.
Indeed, the insight into the percolation problem facilitates the understanding of phase transition and critical 
phenomena that lies at the heart of the modern development of statistical physics.

Percolation on disordered lattice is potentially of great interest 
since many real-life phenomena deal with such disordered systems \cite{ref.Sahimi}. In recent decades there has been
a surge of research activities in studying percolation on random and scale-free network because the 
coordination number disorder of these networks is closely tied to many natural and man-made skeleton or medium through which
percolation occurs. For instance, infectious diseases, computer viruses, opinion, rumors, etc 
 spreads usually through networks \cite{ref.Newman_virus,ref.Moore_virus,ref.Cohen_virus,ref.boccaletti_opinion,
ref.mendes_rumor,ref.pastor_rumor}. Besides, flow of fluids usually takes place through porous medium or 
through rocks and hence the architecture of the skeleton is anything but regular \citep{ref.Gennes,ref.Larson}. 
In fact, transport of fluid through multifractal porous media such as sedimentary strata and in oil reservoir is of 
great interest in geological systems \cite{ref.geology_1, ref.geology_2}.  
In this rapid communication, we investigate percolation on weighted planar stochastic
lattice (WPSL). One of us recently has shown that its coordination number distribution follows a power-law
and its size distribution can be best described as multifractal  \citep{ref.Hassan}. In contrast,
scale-free networks too have power-law coordination number distribution but nodes or sites in the scale-free networks
are neither embedded on spatial positions nor have edges or surfaces. Our goal is to find how the two aspects, multifractality and power-law coordination number distribution, leave their signature in 
the percolation processes.  Classification of percolation into universality 
classes depending on the common sets of critical exponents has been of wide interest. 
It is well-known and widely accepted that the values of the critical exponents depend only on the dimension of the lattice 
and independent of its detailed structure and of the type of percolation, namely site or bond percolation.
It is note worthy to mention 
that when a planar lattice is only multifractal but its dual is not a scale-free network, the resulting percolation still belongs to the same universality as the one for regular planar 
lattice \cite{ref.multifractal}. 
We report for the first time that all the exponents for WPSL are completely different from the existing known values for the planar
lattice ($d=2$) and hence the percolation on WPSL belongs to a new universality class.

\begin{figure}
\includegraphics[width=7.5cm,height=7.0cm,clip=true]{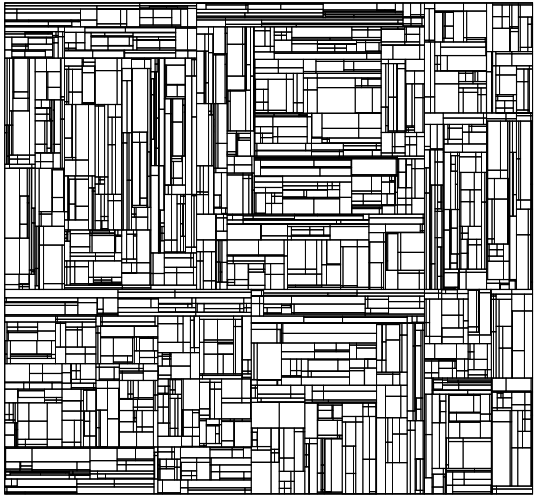}
\caption{A snapshot of the weighted stochastic lattice.
}
\label{fig1}
\end{figure}%

The construction process of the WPSL starts with an initiator, say a square of unit
area. The generator is then defined as the one that divides randomly first the initiator into four smaller blocks.
Thereafter at each step the generator is applied to only one block, by picking it preferentially with respect to areas, and
divides them randomly into four smaller blocks. Each step of the division process
is defined as one time unit. It is thus a process of partitioning an initiator by random sequential division
that creates contiguous blocks of different sizes. The number of blocks $N$ after time $t$ therefore is $N=1+3t$ and hence it 
grows albeit the sum of the areas of all the  blocks remains the same. Thus, the  number of blocks
$N$ increases with time at the expense of the size of the blocks. Indeed, the mean cell area decreases with $N$ like $\langle a\rangle \sim N^{-1}$ or with $t$ like $\langle a\rangle \sim t^{-1}$ 
since $\langle a\rangle=a/N$ where the total area $a$ of all
the blocks is always equal to one. We therefore need to emphasize two things here. First, we
can define the size of the side $L$ of the WPSL as $t^{1/2}$ like we do for square lattice $L=N^{1/2}$. Second, 
to make the cells of the WPSL have the same size, in the statistical sense, as we increase $N$, we have to scale up the cell sizes by a factor of $t$. Else, we just need to multiply each 
quantity we measure by a factor $t$ to compensate 
the decreasing size of the blocks.

Perhaps, the construction process of WPSL is trivially simple but its various properties are far from simple. Firstly, it evolves following several non-trivial conservation laws, 
namely $\sum_i^N x_i^{n-1} y_i^{4/n-1}$ is independent of time or size of the lattice
$\forall \ n$, where $x_i$ and $y_i$ are the length and width of the $i$th block. Secondly, its dual, obtained by replacing each block with a node at its center and common 
border between blocks with an edge joining the two vertices, emerges as a scale-free network \citep{ref.Hassan}. 
Thirdly, if one considers that the $i$th block is populated with  probability $p_i\sim x_i^3 \ {\rm or} \ y_i^3$  
then the $q$th moment of $p_i$ can be shown to exhibit power-law $Z_q(\delta)\sim \delta^{-\tau(q)}$ where $\delta$ is the square root of the mean block area and 
\begin{equation}
\label{massexponent}
\tau(q)=\sqrt{9q^2+16}-(3q+2).
\end{equation} 
Note that $\tau(0)=2$ is the dimension  of the WPSL and $\tau(1)=0$ follows from the normalization of the probabilities $\sum p_i=1$ \cite{ref.multifractal_1}. 
The Legendre transform of $\tau(q)$ on the other hand gives the multifractal spectrum $f(\alpha)$ where the exponent $\alpha$ is the negative derive of $\tau(q)$ with respect to $q$. 
Yet another features of the WPSL is that it 
emerges through evolution and the area size distribution function of its cells exhibits dynamic scaling \cite{ref.hassan_dayeen}.

To study percolation on the WPSL we employ Ziff-Newman algorithm \citep{ref.Ziff} in which all the labeled sites or cells 
$i=1,2,3,..., (1+3t)$ are first randomized and arranged in an order in which the sites 
will be occupied. The good thing about this algorithm is that we can create percolation
states consisting of $n+1$ occupied sites simply by occupying one more site 
to its immediate past percolation state consisting of $n$ occupied sites. Each time thereafter we occupy a site,
it may happen that either an isolated cluster is formed or 
a group of contiguous sites linked by common border may get bigger either by agglomeration or by coagulation.
We keep track of the number of clusters and their sizes
as a function of $n$ {\it vis-a-vis} the occupation probability $p=n/(1+3t)$. In fact, the product of  the number of 
occupied sites $n$ and the mean area $1/(1+3t)$ is equal to the mean area of all the occupied sites $\langle a(n)\rangle$ which
is equal to $1$ when all the sites are occupied i.e., when $n=N$. 
In our simulation we use periodic boundary condition where the lattice is viewed as a torus, thus without edge or surface,
where sites are randomly occupied with probability $p$.

In percolation, one of the primary objectives is to find the occupation probability $p_c$
at which a cluster of contiguous occupied sites span the entire lattice, either horizontally or vertically, 
for the first time. 
Of course, the occupation probability at which it occurs at each independent 
realization on finite size lattice will not be the same. In reality, we can get spanning
even at very much less than $p_c$ or not get it even at a much higher $p$ than $p_c$.
This is exactly why the percolation theory is a part of statistical physics. 
One way of dealing with this is to use the idea of spanning probability $W(p)$ \citep{ref.Ziff_1}.  Consider that we have performed $m$ independent
realizations and for each realization we check exactly at
what value of $p=n/N$ there appears a cluster that connects the two opposite ends either horizontally
or vertically, whichever come first. 
The spanning probability $W(p)$ is the probability of occurrence of spanning cluster. It is obtained by
finding the relative frequency of occurrence of spanning cluster out of $m$ independent realizations. 
The plot in  Fig. (\ref{fig2}), shows $W(p)$ as a function of $p$ for three different lattice sizes. It thus represents the probability of 
finding a spanning cluster at occupation probability $p$ for a fixed lattice size.  One interesting
point is that all the three plots meet at one particular
point. It has a special significance as it means that if we could have data for infinitely large lattice the resulting 
plot would also cross at the same meeting point. This meeting point is actually the percolation threshold $p_c=0.526846$ for
the WPSL.

\begin{figure}
\label{fig2}
\includegraphics[width=5.00cm,height=8.5cm,clip=true,angle=-90]{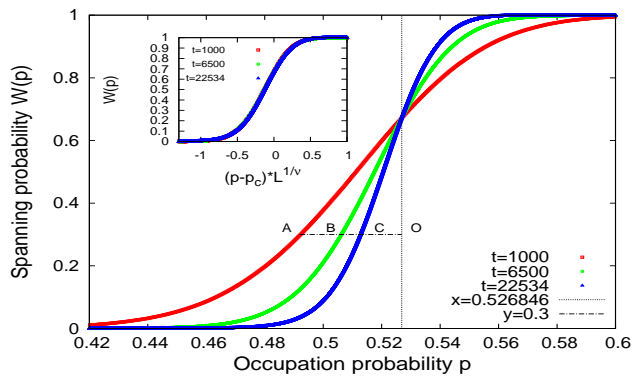}
\caption{Plots of spanning probability $W(p)$ vs  occupation probability $p$ for different lattice size. The vertical is drawn at $p_c=0.526846$. In the inset, 
we plot  $W(p)$ vs $(p-p_c)L^{1/\nu}$, where $\nu=5/3$ and find an excellent data collapse.
\label{fig2}
}
\end{figure}

A careful look at the plots of Fig. (\ref{fig2}) we find that if we increase $L$ then a given fixed value of $W$ is obtained at increasingly higher value of $p$ for $p<p_c$. 
To quantify this 
we draw a horizontal line, for instance at $W(p)=0.3$, and a vertical line passing through the $p_c$ value. Say, the
horizontal line intersects all the three curves and the vertical line for different $L$ at A, B, C and at O. 
We find that the distance $OA, OB, OC$ etc which represents $(p_c-p)$ and plot them in the $\log$-$\log$ scale as a function
of $t$. The resulting plot gives a straight line with slope $0.2966\pm 0.0055$. Using $L\sim t^{1/2}$ we can write
\begin{equation}
\label{eq:3}
(p_c-p) \sim L^{-1/\nu},
\end{equation}
where $1/\nu \sim 0.6$ or $\nu=5/3$. This is different and quite a bit higher than the known value $\nu=4/3$ for all planar lattices. For consistency check, one can now
plot the $p$ values at $A, B, C$ etc versus $L^{-1/\nu}$. The intercept of the resulting linear fit gives the desired $p_c$ value 
and hence this offers an alternative method of measuring $p_c$. The quantity 
$(p_c-p)L^{1/\nu}$ is a dimensionless quantity, according to Eq. (\ref{eq:3}), in the sense that for a given value of $W$ as $L\rightarrow \infty$ the value of $(p_c-p)\rightarrow 0$ 
such that the numerical value of $(p_c-p)L^{1/\nu}$ remains invariant regardless of the lattice
size $L$. We now plot $w(p)$ as a function of $(p_c-p)L^{1/\nu}$, see the inset of Fig. (\ref{fig2}), and find that all the distinct curves of
Fig. (\ref{fig3}) collapse onto a single universal curve. It implies, according to finite size scaling hypothesis, that
\begin{equation}
W(p)\sim L^\eta \phi\Big( (p-p_c)L^{1/\nu}\Big ), 
\end{equation}
with exponent $\eta=0$ where $\phi$ is the scaling function \cite{ref.saberi}. It states that the spanning probability $W$ itself is a dimensionless quantity provided it is measured in the scaled variable  
$(p_c-p)L^{1/\nu}$ \cite{ref.barenblatt}. 
It also means that the spanning probability for infinite lattice size would be
like a step function around $p_c$.

\begin{figure}
\label{fig3}
\includegraphics[width=5.00cm,height=8.5cm,clip=true,angle=-90]{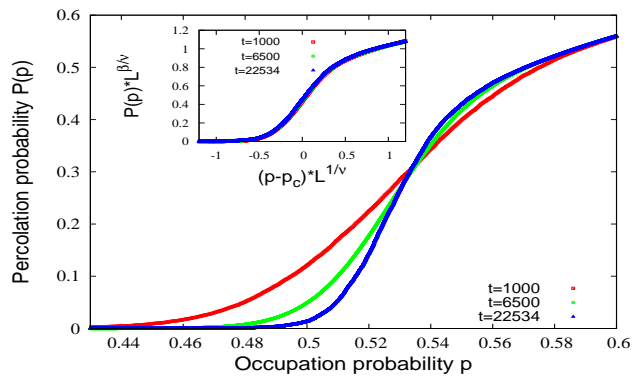}
\caption{Plots of percolation probability $P(p)$ vs $p$ for three different size of the WPSL. In the
inset we plot the same data but in the scaled variables $PL^{\beta/\nu}$ and $(p_c-p)L^{1/\nu}$ and  find an excellent data-collapse.
\label{fig3}
}
\end{figure}

It is well-known that like Ising model percolation too display a continuous phase
transition and hence like magnetization of the Ising model there must be an order parameter for the percolation
model too. The fact is that not all the occupied sites 
belong to spanning cluster. We thus can define the percolation probability $P$ which must be zero 
below $p_c$ and should increase continuously beyond $p_c$ - a characteristic feature for order parameter.  
We define it as the ratio of the area of the spanning cluster
$A_{{\rm span}}t$ to the total area of the lattice $at$ and hence
$P(p)= A_{{\rm span}}$
since the the total area of the lattice is always equal to one. 
Unlike $W(p)$ vs $p$ the distinct curves of the the $P(p)$ vs $p$ plots, see Fig. (\ref{fig3}), for different size do not meet at one unique value,
namely at $p_c$ which we can only appreciate if we zoom in. Nevertheless,
following the same procedure we once again
find  $(p_c-p)\sim L^{-1/\nu}$ with the same $\nu$ value. Like for $W(p)$ if we plot $P$ as a function $(p_c-p)L^{1/\nu}$
we do not get data collapse as before, instead we see that for a given value of $(p_c-p)L^{1/\nu}$ the $P$ value 
decreases with lattice size $L$ following a power-law 
\begin{equation}
\label{eq:4}
P\sim L^{-\beta/\nu},
\end{equation}
where $\beta/\nu=0.135\pm 0.0076$. It implies that for a given value of $(p_c-p)L^{1/\nu}$ the numerical value of $PL^{\beta/\nu}$ must remain invariant regardless of the lattice size of $L$. That is, 
if we now plot $PL^{\beta/\nu}$ vs $(p_c-p)L^{1/\nu}$ all the distinct plots of $P$ vs $p$ should collapse into a single universal
curve. Indeed, such data-collapse is shown in the inset of Fig. (\ref{fig3}) which implies that percolation probability $P$
exhibits finite-size scaling
\begin{equation}
P(p_c-p,L)\sim L^{-a}\phi\Big ((p_c-p)L^{1/\nu}\Big ).
\end{equation}
Now, eliminating $L$ in favor of $p_c-p$ in Eq. (\ref{eq:4}) we get
\begin{equation}
\label{eq:5}
P\sim (p_c-p)^\beta,
\end{equation}
where $\beta\sim 0.225$ or $\beta=9/40$ for WPSL whereas $\beta=5/36$ for all other known planar lattices.




\begin{figure}
\label{fig4}
\includegraphics[width=5.00cm,height=8.5cm,clip=true,angle=-90]{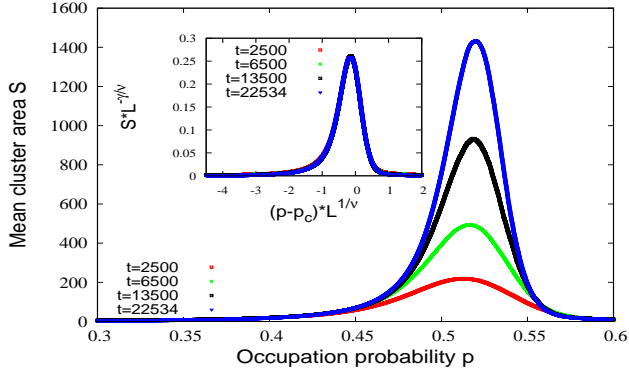}
\caption{Mean cluster area S(p) as function of occupation probability $p$ with different lattice size $L$. In the
inset we plot the scaled variables $SL^{-\gamma/\nu}$ vs $(p_c-p)L^{1/\nu}$ and find that data for different system sizes are
well collapsed in a single universal curve.
\label{fig4}
}
\end{figure}

Percolation is all about clusters and hence the cluster size distribution function $n_s(p)$ plays a central
role in the description of the percolation theory.  It is defined as the number of clusters of size $s$ per site.
The quantity $sn_s(p)$ therefore is the probability that an arbitrary site belongs to a cluster of $s$ sites 
and $\sum_{s=1} sn_s$ is probability that an arbitrary site belongs to a cluster of any size which is in fact equal to
$p$.
The mean cluster size $S(p)$ therefore is given by
\begin{equation}
\label{eq:6}
S(p)=\sum_s sf_s={{\sum_s s^2n_s}\over{\sum_s sn_s}},
\end{equation}
where the sum is over the finite clusters only. In the case of percolation on the WPSL, we regard $s$ as
the cluster area. It is important to mention that each time we evaluate the ratio of the second and the first moment of
of $n_s$ we also have to multiply the result by $t$, the time at which the snapshot of the lattice is taken, to compensate the decreasing block size with increasing block number $N$. 
The mean cluster size therefore is
$S={{1}\over{p}}\sum_s s^2n_s \times t$ where $\sum_s sn_s=p$ is the sum of the areas of all the clusters. Note that the spanning cluster is excluded 
from both the sums of Eq. (\ref{eq:6}). 
In Fig. (\ref{fig4}) we plot $S(p)$ as a function of $p$ for different lattice sizes $L$. We observe that there are two main effects as we
increase the lattice size. First, we see that the mean cluster area
always increases as we increase the occupation probability. However, as the $p$ value approaches to $p_c$, we find
that the peak height grows profoundly with $L$.

The increase of the peak height can be quantified by plotting 
these heights as a function of $L$ in the $\log$-$\log$ scale and find
\begin{equation}
\label{eq:7}
S\sim L^{\gamma/\nu},
\end{equation}
where $\gamma/\nu=1.73\pm 0.006321$. A careful observation reveals that there is also a shift in the $p$ value at which the peaks occur. 
We find that the magnitude of this shift $(p_c-p)$ becomes smaller with
increasing $L$ following a power-law $(p_c-p)\sim L^{-1/\nu}$. 
 We now plot the same
data in Fig (\ref{fig4}) by measuring the mean cluster area
$S$ in unit of $L^b$ and $(p_c-p)$ in unit of $L^{-1/\nu}$ respectively and find that all the distinct plots of
 $S$ vs $p$ collapse into one universal curve, see the inset of the same figure. It again implies that the mean cluster area too exhibits finite-size scaling
\begin{equation}
\label{eq:8}
S \sim L^{b}\phi \Big ((p_c-p)L^{1/\nu}\Big ).
\end{equation}
Eliminating $L$ from Eq. (\ref{eq:7}) in favor of $(p_c-p)$ using $(p_c-p)\sim L^{-1/\nu}$ we find that the mean cluster
area diverges 
\begin{equation}
\label{eq:9}
S\sim (p_c-p)^{-\gamma},
\end{equation}
where $\gamma=2.883$ which we can approximately write $\gamma=173/60$. In contrast, $\gamma=43/18$ for all other planar lattices.

\begin{figure}
\centering
\label{fig:ab}
\subfloat[]
{
\includegraphics[height=1.6in, width=1.6in, clip=true,angle=-90]{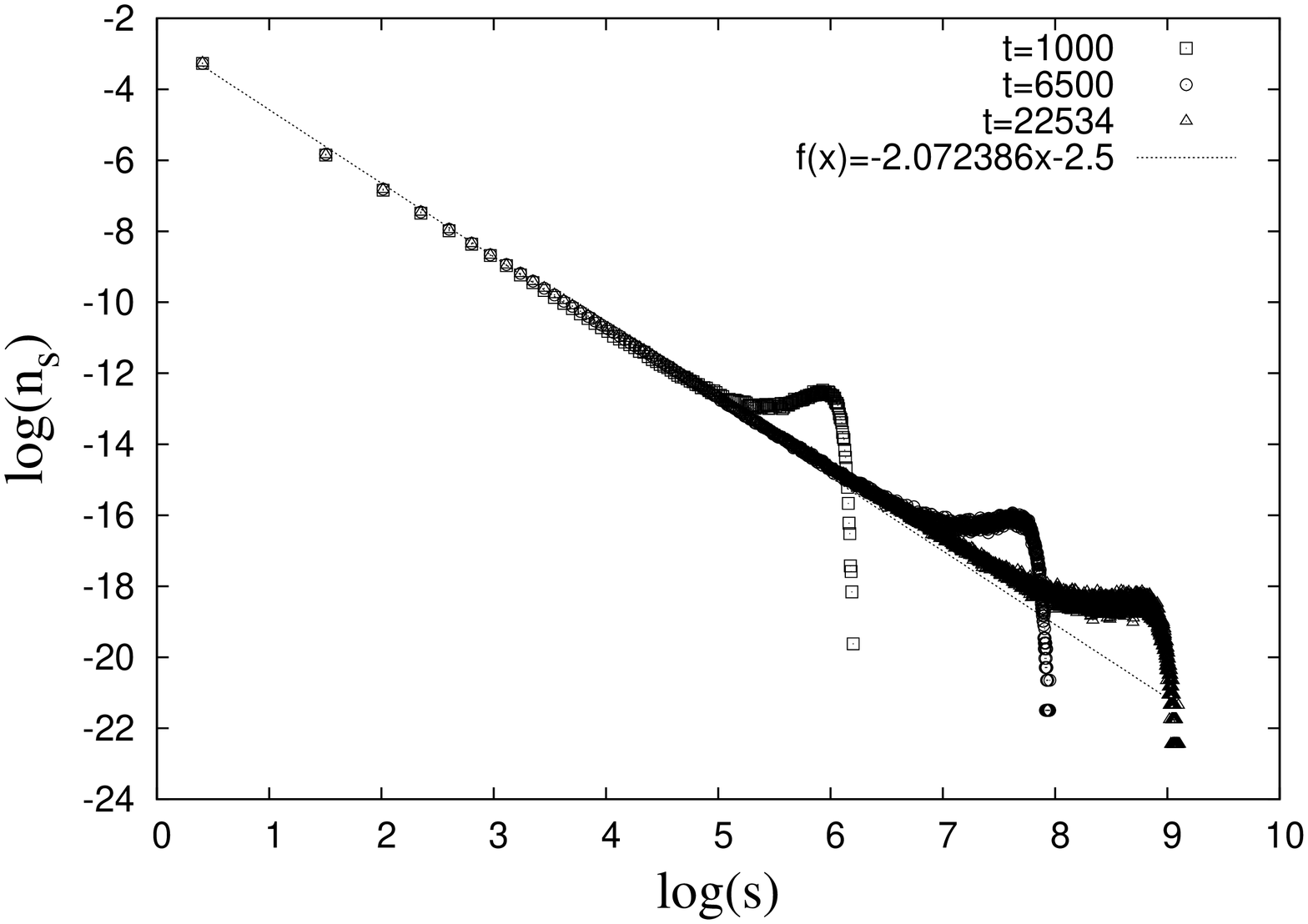}
\label{fig:a}
}
\subfloat[]
{
\includegraphics[height=1.6in, width=1.6in, clip=true,angle=-90]{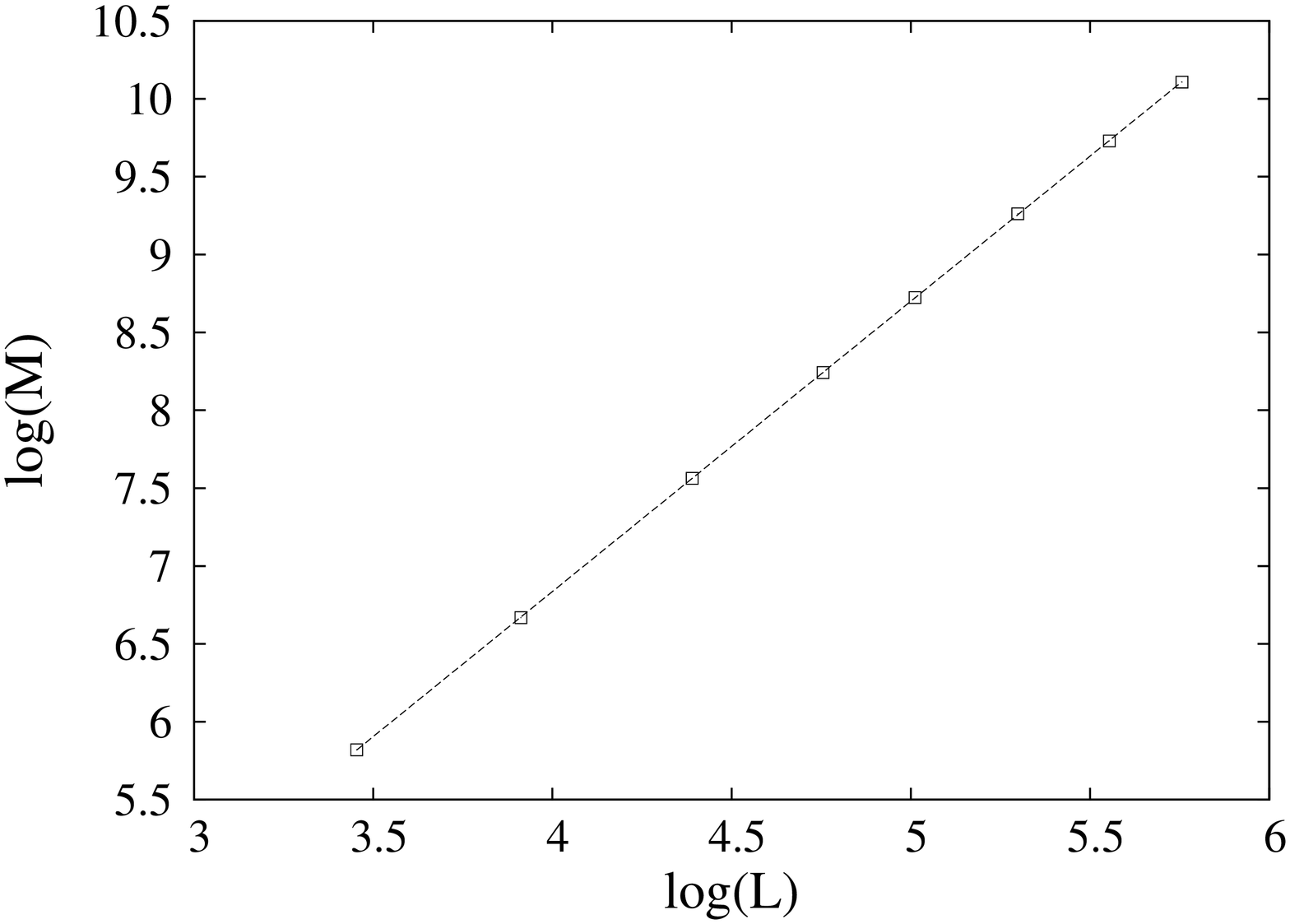}
\label{fig:b}
}
\caption{(a) The plot of $\log(n_s)$ vs $\log(s)$  for different lattice size $L$.
(b) The double-logarithmic plot of the size of the spanning cluster $M$ against the lattice size $L$. 
} 
\label{fig:ab}
\end{figure}

We can also obtain the exponent $\tau$ by plotting the cluster area distribution function $n_s(p_c)$ at $p_c$. We
plot it in the log-log scale and find a straight line except near the tail. However, we also observe that 
as the lattice size increases the extent up to which we get a straight line having the same slope increases. It implies
that if we performed on WPSL of infinitely large size we would have a perfect straight line obeying
$n_s(p_c)\sim s^{-\tau}$
with $\tau=2.0724$ which is less than its value for all other known planar lattices $\tau=187/91$. We already know that mean
cluster area $S\rightarrow \infty$ as $p\rightarrow p_c$. According to Eq. (\ref{eq:6}), $S$ can only diverge
if its numerator diverges. Generally, we know that $\sum_{s=1}^\infty s^\alpha$ converges if $\alpha<-1$ and diverges if 
$\alpha\geq -1$. Applying it into both numerator and denominator of Eq. (\ref{eq:6}) at $p_c$ gives a bound that
$2<\tau<3$. Assuming
\begin{equation}
n_s(p)\sim s^{-\tau}e^{-s/s_\xi},
\end{equation}
and using it in Eq. (\ref{eq:6}) and taking continuum limit gives
\begin{equation}
S\sim s_\xi^{3-\tau}.
\end{equation}
We know that $s_\xi$ diverges like $(p_c-p)^{1/\sigma}$ where $\sigma=1/(\nu d_f)$ and hence comparing it with Eq. (\ref{eq:9})
we get
\begin{equation}
\label{eq:tau}
\tau=3-\gamma \sigma.
\end{equation}
Note that the ramified nature of the spanning cluster at $p_c$ is reminiscent of fractal. Indeed, we find that the 
the fractal dimension $d_f$ of the spanning cluster can be obtained by finding 
the gradient of the plot of the size of the spanning cluster $M$ as a function of lattice size $L$ in the log-log scale (see Fig (\ref{fig:b}). We find $d_f=1.865$ which 
can also be written as $d_f=373/200$ for WPSL and that for  square, triangular, honeycomb,
Voronoi lattices is $d_f=93/48$. Using the value of $\gamma$ and $\sigma$ in Eq. (\ref{eq:tau}) we get 
$\tau=2.0724$ which we can approximately write as $773/373$. This is consistent with what we found from the slope of $\log[n_s(p_c)]$ vs $\log[s]$ plot shown in Fig. (\ref{fig:a}).

\begin{center}
    \begin{tabular}{| l | l | l |}
    \hline
    Exponents & regular planar lattice & WPSL \\ \hline
    $\nu$ & 4/3 & 5/3  \\ \hline
    $\beta$ & 5/36 & 9/40  \\ \hline
    $\gamma$ & 43/18 & 173/60  \\ \hline
   $ \tau$ & 187/91 & 773/373 \\ \hline
$d_f$ & 91/48 & 373/200 \\

    \hline
    \end{tabular}
\end{center}

To summarize, we have studied percolation on a scale-free multifractal planar lattice. We obtained the $p_c$ value and
the characteristic exponents $\nu, \beta,\gamma, \tau, \sigma$ and $d_f$ which characterize the percolation transition.  
Note that it is the sudden onset of a spanning cluster at the threshold $p_c$ which is 
accompanied by discontinuity or divergence of 
some observable quantities at the threshold make the percolation transition a critical phenomena. One of the most interesting and useful
aspects of percolation theory so far known is that the values of the various exponents depend only on the dimensionality of the 
lattice as they are found independent of
the type of lattice (e.g., hexagonal, triangular or square, etc.) and the type of percolation (site or
bond).  This central property of percolation theory is known as "universality". 
Recently, Corso {\it et al} performed percolation on 
a particular mutifractal planar lattice whose coordination number distribution is, however, not scale-free like
WPSL and still they found the exponents as for all the planar regular lattices \cite{ref.multifractal}. 
Thus the most expected result would be to find a different value for $p_c$ value as its coordination number
distribution is totally different than any known planar lattice. However, finding
a complete different set of values, see the table, for all the characteristic exponents was not expected since WPSL too a planar
lattice. Interestingly, like existing values for regular planar lattices, the exponents of the values for WPSL too satisfy the scaling relations $\beta=\nu(d- d_f)$, $\gamma=\nu(2d_f-d)$, $\tau=1+d/d_f$. 
We can this conclude that percolation on WPSL belongs to a new universality class. 
It would be interesting to check the role of the exponents $\gamma$ of the power-law coordination number distribution in the classification of universality classes. We intend to do it in our future endeavour.

We gratefully acknowledge A. A. Saberi for critical reading of the manuscript and for valuable comments.


\begin{thebibliography}{99}
\bibitem{ref.Stauffer} D. Stauffer and A. Aharony, {\it Introduction to Percolation Theory} (Taylor $\&$ Francis, London, 1994).
\bibitem{ref.bunde} {\it Fractals and Disordered Systems} Edited by A. Bunde and S. Havlin  
(New York, NY, Springer Verlag, 1996).
\bibitem{ref.Stanley} H. E. Stanley, {\it Introduction to Phase Transitions and Critical Phenomena} (Oxford University Press, Oxford and New York 1971).
\bibitem{ref.Binney} J.  J.  Binney,  N.  J.  Dowrick,  A.  J.  Fisher,  and  M.  E.  J.  Newman,  {\it The  Theory  of
Critical  Phenomena}  (Oxford University Press, New York, 1992).
\bibitem{ref.Sahimi} M. Sahimi, {\it Applications of Percolation Theory} (Taylor $\&$ Francis, London, 1994).
\bibitem{ref.Newman_virus} M.E.J. Newman and D.J. Watts, Phys. Rev. E 60, 7332. (1999).
\bibitem{ref.Moore_virus}  C. Moore and M.E.J. Newman, Phys. Rev. E62, 7059. (2000).
\bibitem{ref.Cohen_virus} R. Cohen, K. Erez, D. ben-Avraham, and S. Havlin, Phys. Rev. Lett. 85, 4626-4628 (2000).
\bibitem{ref.boccaletti_opinion} Boccaletti, S., Latora, V., Moreno, Y., Chavez, M., Hwang, D.: Phys. Rep.
424, 175 (2006).
\bibitem{ref.mendes_rumor}5. S. N. Dorogovtsev, J. F. F. Mendes, {\it Evolution of Networks} (Oxford University Press, Oxford 2003).
\bibitem{ref.pastor_rumor} R. Pastor-Satorras and A. Vespignani, {\it Evolution and Structure of the Internet: A Statistical Physics Approach} (Cambridge University Press, Cambridge 2004).
\bibitem{ref.Gennes} P.G. de Gennes and E. Guyon, J. de Mecanique 3, 403 (1978).
\bibitem{ref.Larson} R. G. Larson, L. E. Scriven, and H. T. Davis, Chem. Eng. Sci. 15, 57 (1981).
\bibitem{ref.geology_1} J. Muller, Ann. Geophys. {\bf 11} 525 ?(1993?).
\bibitem{ref.geology_2} P. N. Khue, O. Fluseby, A. Saucier, and J. Muller, J. Phys.:
Condens. Matter {\bf 14} 2347 ?(2002?).
\bibitem{ref.Hassan} M. K. Hassan, M. Z. Hassan, and N. I. Pavel, New Journal of Physics {\bf 12}  093045 (2010); {\it ibid} J. Phys: Conf. Ser, {\bf 297} 012010 (2011).
\bibitem{ref.multifractal} G. Corso, J. E. Freitas, L. S. Lucena, and R. F. Soares, Phys. Rev E {\bf 69} 066135 (2004).
\bibitem{ref.multifractal_1} J. Feder,  {\it Fractals} (Plenum, New York,  1988).
\bibitem{ref.hassan_dayeen} F. R. Dayeen and M. K. Hassan, arXiv:1409.7928 [cond-mat].
\bibitem{ref.Ziff} M. E. J. Newman  and R. M. Ziff. Phys. Rev. Lett. {\bf 85} 4104 (2000); {\it ibid} Phys. Rev. E {\bf 64} 016706 (2001).
\bibitem{ref.Ziff_1} R. M. Ziff, Phys. Rev. Lett., 69:2670, 1992.
\bibitem{ref.saberi} A. A. Saberi, Appl. Phys. Lett. {\bf 97} 154102 (2010).
\bibitem{ref.barenblatt} G. I. Barenblatt, {\it Scaling, Self-similarity, and Intermediate Asymptotics} (Cmpridge University Press, 1996).







\end{thebibliography}
\end{document}